\documentclass[prl,twocolumn,letterpaper,showpacs]{revtex4}

\usepackage{graphics}
\usepackage{epsfig}
\usepackage{amssymb}

\begin{document} 

\title{Vortex arrays in a rotating superfluid Fermi gas}

\author{David~L.~Feder} 
\affiliation{Department of Physics and Astronomy, University of Calgary, 
Calgary, Alberta, Canada T2N 1N4}

\date{\today}

\begin{abstract} 

The behavior of a dilute two-component superfluid Fermi gas subjected to 
rotation is investigated within the context of a weak-coupling BCS theory. 
The microscopic properties at finite temperature are obtained by iterating 
the Bogoliubov-de Gennes equations to self-consistency. In the model, alkali 
atoms are strongly confined in quasi-two-dimensional traps produced by 
a deep one-dimensional optical lattice. The lattice depth significantly 
enhances the critical transition temperature and the critical rotation 
frequency at which the superfluidity ceases. As the rotation frequency 
increases, the triangular vortex arrays become increasingly irregular, 
indicating a quantum melting transition.

\end{abstract}


\maketitle

Great experimental strides have been made over the past year in the goal to 
form a BCS-like superfluid state with ultracold neutral Fermi gases. Through
the use of magnetic field-induced Feshbach 
resonances~\cite{Feshbach, Stwalley,Tiesinga}, strong interactions between 
quantum degenerate fermions in two different hyperfine states have been 
induced~\cite{Regal1,Bourdel,Gupta,OHara1}. Molecules result when the 
resulting interactions are 
repulsive~\cite{Regal2,Cubizolles,Jochim,Strecker,Zwierlein1}; 
though Cooper pairing at high temperatures
(within an order of magnitude of the degeneracy temperature $T_F$) is widely 
expected for attractive interactions near the Feshbach 
resonance\cite{Kokkelmans,Ohashi}, the state actually produced in recent 
experiments~\cite{Greiner,Zwierlein2} remains to be fully elucidated. 

Another proposed approach to high-temperature superfluidity in these systems 
is to confine the gases in optical lattices~\cite{Hofstetter,Modugno}. The 
flattening of the energy bands as the lattice depth increases ensures that a 
large number of atoms participate in the pairing even for relatively weak 
interactions. A deep one-dimensional (1D) lattice corresponds to an array of 
quasi-2D traps, whose large effective axial confinement can enhance the 
interaction strength sufficiently~\cite{Petrov1} to raise the superfluid 
transition temperature $T_c$ to within $10\%$ of $T_F$~\cite{Petrov2}. 

Rotating the superfluid around the optical lattice axis should yield vortex 
arrays akin to those observed recently in trapped Bose-Einstein 
condensates~\cite{Coddington}.  Few clear signatures of superfluidity exist 
for dilute Fermi gases~\cite{Bruun1}; the first evidence has been obtained only 
very recently~\cite{Kinast,Bartenstein,Chin}. The experimental observation of 
quantized vortices would be a `smoking gun' for the presence of superfluidity 
in the system, though significant depletion of particles in the vortex cores is 
expected only in the strong-coupling limit~\cite{Bulgac1}. Vortices in 
the array of weakly coupled two-dimensional (2D) traps serve as a clean model 
of the pancake vortices in strongly layered high-$T_c$ superconductors such as 
Bi$_2$Sr$_2$CaCu$_2$O$_{8+\delta}$ (BSCCO)~\cite{Clem}. Furthermore, 
because a small fraction of the total number of atoms reside in any individual 
lattice site, one might expect to readily access the quantum Hall limit at high 
rotation frequencies where the number of vortices becomes comparable to the 
number of atoms~\cite{Cooper,Popp}. 

To my knowledge, the calculations presented below constitute the first 
microscopic determination of vortex arrays in inhomogeneous Fermi superfluids. 
The central results of the work are: (1) the critical rotation frequency for
the cessation of superfluidity (the rotational analogue of the critical 
velocity $v_c$ in superconductors) scales with $T_c$; (2) the vortex 
cores for superfluids in the BCS limit that are confined in 1D optical lattices
should be directly visible experimentally at low temperatures; and (3) the 
vortex lattice is expected to melt at zero temperature at high rotation 
frequencies approaching the expected quantum Hall transition.

The calculations were inspired by experiments with attractive Fermi gases 
confined in optical traps~\cite{OHara1,Zwierlein2}. The experiments were 
performed at magnetic fields above but near the Feshbach resonance at 
860~G~\cite{OHara2} at which the $s$-wave scattering length between atoms in 
the hyperfine states $\left|F={1\over 2},m_F=\pm{1\over 2}\right\rangle$ 
(labeled $\sigma=\uparrow,\downarrow$ below) is $a\sim -10^4a_0$
($a_0=0.0529$~nm). With $N=7.5\times 10^4$ atoms in each species~\cite{OHara1}, 
the Thomas-Fermi (TF) approximation $N={1\over 6}{\omega_{\rho}\over\omega_z}
({\mu_{\rm TF}\over\hbar\omega_{\rho}})^3$ can be inverted to yield the TF 
chemical potential $\mu_{\rm TF}\approx 25\hbar\omega_{\rho}$ and the Fermi 
momentum at the trap center $\hbar k_F^0\approx\sqrt{2m\mu_{\rm TF}}$;
one then obtains $k_F^0|a|\gg 1$ which implies very strong coupling.

In contrast, the BCS mean-field approximation is applicable only for
$k_F^0|a|\lesssim 0.6$, beyond which one needs to include pairing 
fluctuations~\cite{Perali}. The BCS transition temperature is given 
approximately by the uniform 3D expression~\cite{Stoof}
$T_c=(8e^{\gamma-2}/\pi)\mu_{\rm TF}\exp[-\pi/(2k_F^0|a|)]$, where 
$\gamma=0.5771\cdots$ is Euler's constant; this does not 
take into account polarization corrections (not considered in the present 
work), which lower $T_c$ by a factor of approximately 
2.2~\cite{Gorkov}. Decreasing $|a|$ such that $k_F^0|a|=0.6$ yields 
$T_c\approx 1.1\hbar\omega_{\rho}\approx 0.06T_F$ which remains too low to 
easily access experimentally.

This value for $T_c$ can be substantially increased if the laser were 
retro-reflected to form a 1D optical lattice, with potential 
$V_{\rm lat}=nE_R\sin^2(2\pi z/\lambda)$ where $n$ is the lattice depth in
units of the recoil energy $E_R={\hbar^2\over 2m}({2\pi\over\lambda})^2$.
Near the center the potential may be approximated as quadratic in $z$, with
the ratio of the effective axial and radial frequencies 
$\alpha\equiv\tilde{\omega}_z/\omega_{\rho}=\sqrt{n}(2\pi d_{\rho}/\lambda)^2
=20-80$ where $d_{\rho}=\sqrt{\hbar/m\omega_{\rho}}$ and the lower and upper 
limits correspond to $\lambda=1064$~nm, $n\approx 0.4E_R$~\cite{Zwierlein2} and $\lambda=10.6~\mu$m, $n\approx 50\,000E_R$~\cite{OHara1}. Neglecting inter-well 
tunneling, all of the atoms occupy the axial oscillator ground state in the 
quasi-2D limit $\mu\ll\hbar\tilde{\omega}_z$.  A straightforward 
calculation~\cite{comment} gives the superfluid transition temperature for a 
uniform quasi-2D system $k_BT_c^{\rm 2D}=(2e^{\gamma}/\pi)\mu_{\rm TF}^{\rm 2D}
\exp(-\sqrt{\pi\over 2}{\tilde{d}_z\over |a|})$ where 
$\tilde{d}_z=\sqrt{\hbar/m\tilde{\omega}_z}$, not including polarization
effects which reduce $T_c^{\rm 2D}$ by a factor of $e$~\cite{Petrov2}. In the
calculations below, $\mu=10\hbar\omega_{\rho}$, $d_{\rho}=1~\mu$m, $a=-2160a_0$
(corresponding to the triplet scattering length that would be obtained for an
external bias field around 2000~G far above the Feshbach resonance), and 
$20\le\alpha\le 40$. Choosing $\alpha=40$, one obtains 
$T_c^{\rm 2D}\approx 0.16~\mu$K$\approx T_F/5$. This significantly improves 
$T_c$ while $k_F^{2D}|a|\approx 0.5$ remains in the weak-coupling limit.

The calculations are based on a BCS mean-field approximation to the full 
2D interaction Hamiltonian~\cite{deGennes}, where the particle density and gap
functions are defined by the thermal averages 
$n_{\sigma}({\mathbf{x}})=\langle\psi^{\dag}_{\sigma}({\mathbf x})
\psi^{\vphantom{\dag}}_{\sigma}({\mathbf x}) \rangle$ and
$\Delta({\mathbf{x}})=-\tilde{g}'\langle\psi_{\uparrow}({\mathbf x})
\psi_{\downarrow}({\mathbf x}) \rangle$, respectively (here and below 
${\bf x}=(x,y)$ corresponds to a 2D vector). The ultraviolet
divergence in the definition of the superfluid gap is regularized using the
pseudopotential method~\cite{Bruun2,Bulgac2}, giving rise to a 
regularized coupling constant $\tilde{g}'$. An equal population of the two 
hyperfine components $N_{\uparrow}=N_{\downarrow}$ is chosen to maximize 
$T_c$~\cite{Stoof}. Diagonalizing the mean-field Hamiltonian in a frame 
rotating around the lattice axis at angular frequency $\Omega$, one obtains 
the Bogoliubov-de Gennes (BdG) equations~\cite{deGennes} 
\begin{eqnarray}
\left[\begin{array}{cc}{\mathcal{H}}-\mu&\Delta({\mathbf{x}})\\
\Delta^*({\mathbf{x}})&-({\mathcal{H}}-\mu)^*
\end{array}
\right]\left[\begin{array}{c}u_n({\mathbf{x}})\\v_n({\mathbf{x}})\end{array}
\right]=
E_n\left[\begin{array}{c}u_n({\mathbf{x}})\\v_n({\mathbf{x}})\end{array}
\right].
\label{BdG}
\end{eqnarray}
Here ${\mathcal{H}}=-\frac{\hbar^2}{2m}\nabla_{\bf x}^2
+{1\over 2}m\omega_{\rho}^2\left(x^2+y^2\right)+g'n_\sigma({\bf x})-\Omega L_z$,
where $L_z=i\hbar(y\partial_x-x\partial_y)$. The density and gap functions are
$n_{\sigma}({\bf x})=\sum_n\left[|u_n|^2f(E_n)+|v_n|^2(1-f(E_n))\right]$
and $\Delta({\bf x})=-\tilde{g}'\sum_nu_nv_n^*(1-2f(E_n))$, respectively,
and must be iterated to self-consistency together with the quasiparticle
amplitudes $u_n$ and $v_n$ appearing in Eqs.~(\ref{BdG}) to obtain equilibrium
solutions; the sums run over positive 
$E_n$ and $f(E_n)=(e^{E_n/k_B T} +1)^{-1}$.

The BdG matrix was evaluated numerically using a DVR based on Hermite 
polynomials with up to 100 functions/points each in $x$ and $y$. The 
eigenvectors for energies up to $E_n({\rm max})$ were obtained using the 
routine {\tt pzheevx} on a 32-processor Xeon cluster. For the largest grid, 
each diagonalization took approximately 50 minutes. The procedure was deemed 
converged when the magnitudes of $\Delta$ and $n_{\sigma}$ at successive 
iterations were smaller than some predefined tolerance; for large $\Omega$ as 
many as 3000 iterations were required to verify that the ground state had been
obtained. The sums over $E_n$ were found to fully converge for 
$E({\rm max})\sim3\mu$. $N_{\sigma}$ was on the order of 3000 for 
$\Omega=0.98\omega_{\rho}$ for $\alpha=40$.

\begin{figure}[t]
\epsfig{file=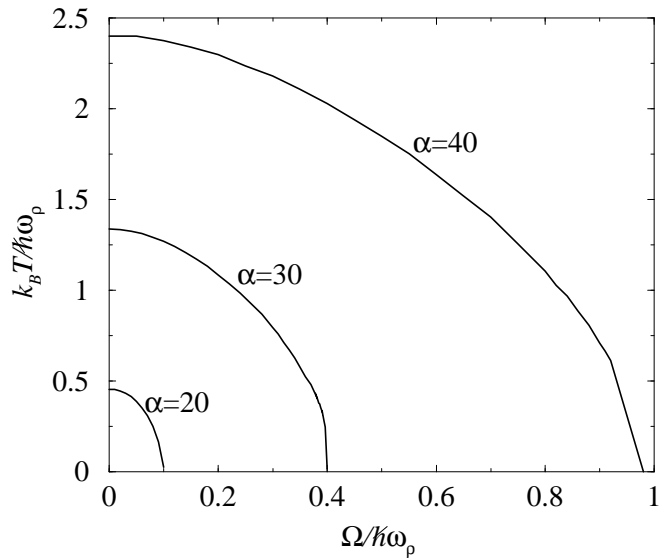,width=\columnwidth,angle=0}
\caption{$(T,\Omega)$ phase diagram for the quasi-2D system as a function of
$\alpha=\tilde{\omega}_z/\omega_{\rho}$. Regions to the left (right) of a 
given curve correspond to superfluid (normal) states. Note that 
$\hbar\omega_{\rho}/k_B\approx 80$~nK, so that $T_c(\Omega=0)\approx 200$~nK
for $\alpha=40$.}
\label{phase}
\end{figure}

Because the effective coupling strength $g'\propto\sqrt{\alpha}\propto n^{1/4}$ 
(where $n$ is the lattice depth in recoils) is a parameter adjustable by 
varying the trapping laser intensity, it is useful to explore the $(T,\Omega)$ 
superfluid phase diagram for various $\alpha$, subject to the constraints
$\mu/\hbar\omega_{\rho}<\alpha$ (single axial mode approximation) and 
$k_F^0|a|<1$ (weak coupling). Results for $\alpha=\{20,30,40\}$ are shown in 
Fig.~\ref{phase}; the superfluid transition was obtained assuming cylindrical 
symmetry (when $\Delta=0$ there are no vortices) with random points checked 
using the full 2D code. The numerical value of $T_c$ for $\Omega=0$ is 
comparable to the TF value of $T_c^{\rm 2D}$ given above, in spite of the 
small total number of atoms and the inhomogeneous density. Superfluidity also
ceases at zero temperature at a critical angular frequency $\Omega_c$. This
is entirely due to the breaking of Cooper pairs by the quasiparticles'
rotational motion, and is the neutral analog of the critical velocity $v_c$ in 
uniform superconductors subjected to a voltage drop~\cite{deGennes}. In the 
latter case, $v_c\propto\propto\Delta(T=0)\propto T_c$; the numerical results 
clearly indicate that while the critical frequency is proportional to $T_c$, it 
is much smaller than $\omega_{\rho}$ unless $T_c$ is enhanced by the trap 
anisotropy.

\begin{figure}[t]
\epsfig{file=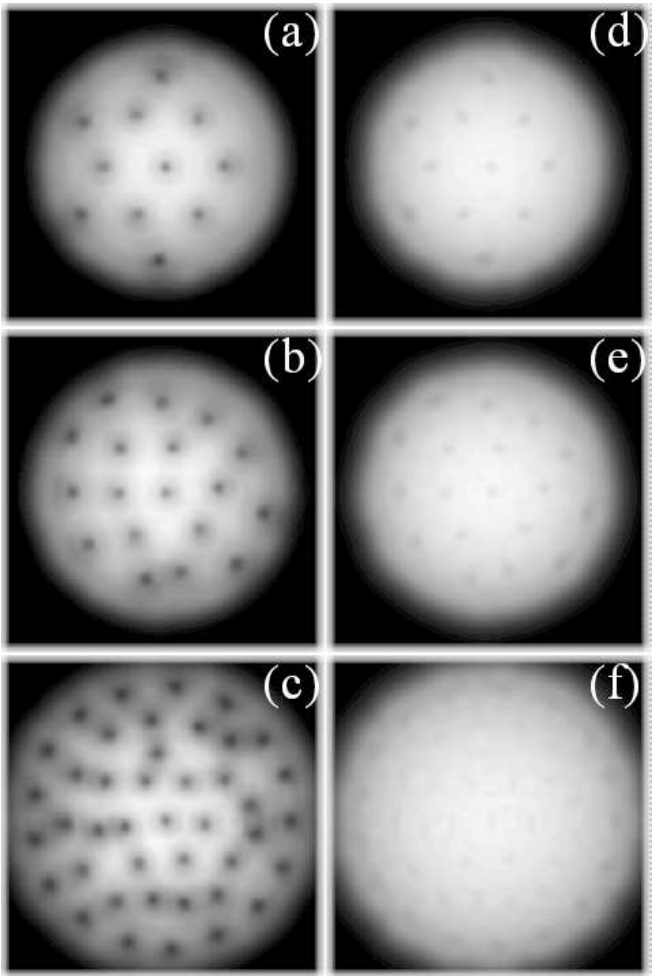,width=\columnwidth,angle=0}
\caption{Density plots for the in-trap gap function $\Delta({\bf x})$ (a-c) 
and number density $n_{\sigma}({\bf x})$ (d-f) for $\alpha=40$ and $T=0$. 
Figures (a,d), (b,e), and (c,f) correspond to $\Omega/\omega_{\rho}=0.4$, 
$0.5$, and $0.7$, respectively. Each box is 16~$\mu$m on a side.}
\label{fig2}
\end{figure}

The immediate question is: can $\Omega_c$ be made sufficiently large that 
one or more vortices can be stabilized? If $\Omega$ is too small, the 
vortex-free state is energetically favored, and for $\Omega\sim\Omega_c$ the 
magnitude and spatial extent of $\Delta$ are too small to support vortices 
whose characteristic size is the local healing length 
$\xi=\hbar^2k_F/m\pi\Delta$.
No vortices were observed for $\alpha=20$, while $\alpha=30$ yielded a small 
number, fewer than 10. As shown in Fig.~\ref{fig2}, the $\alpha=40$ case, with 
$\Omega_c\sim\omega_{\rho}$, is able to support large numbers. For small 
$\Omega$, the vortices distribute themselves into a regular triangular pattern;
however, for increasing angular frequency the arrays become disordered, so 
that by $\Omega=0.7\omega_{\rho}$ the $T=0$ lattice has completely melted. 
These results were independent of the initial guesses for $\Delta$ and 
$n_{\sigma}$ and of the convergence criterion. The vortices tend 
to be found on circles defined by the boundary between orbitals of angular 
momentum $m$ and $m+1$, and concomitantly exhibit considerable azimuthal 
distortion (the effect is most noticeable in the number density).

The zero-temperature melting of a 2D vortex array was suggested some time ago
in the context of the layered high-$T_c$ superconductors without
impurities~\cite{Chudnovsky,Rozhkov}. According to the Lindemann criterion,
melting occurs when the zero-point amplitude of vortex position fluctuations
is some critical fraction $c_L$ of the intervortex separation $\ell_v$; for 
most materials, $c_L\approx 0.1-0.2$. The quasiparticle bound states in the
vortex core account for the vortex fluctuations, and have a spatial extent 
$\sim\xi$. The vortex separation $\ell_v$ for a given $\Omega$ is determined
by matching the angular velocity of the normal-state atoms with the average
circulation $\kappa=h/2m$ of the superfluid vortices, which yields the vortex
density $n_v=2\Omega/\kappa$ or $\ell_v=1/\sqrt{n_v}\approx 2d_{\rho}=2~\mu$m
which depends only weakly on $\Omega$. For $\Omega=\{0.4,\,0.5,\,0.7\}$ shown 
in Fig.~\ref{fig2}(a-c), the
numerics yield $\xi/\ell_v\approx\{0.15,\,0.17,\,0.2\}$, consistent with 
the Lindemann criterion. Indeed, visual inspection of the lowest-energy 
quasiparticle amplitudes $u_n$ and $v_n$ reveals that they are highly localized 
near the vortex cores for $\Omega=0.4$ but begin to overlap those of adjacent
vortices for larger $\Omega$. Furthermore, for systems in the lowest Landau 
level the relation $N/N_v\sim c_L^{-1}$ is expected~\cite{Rozhkov}; for the
values of $\Omega$ considered above one obtains 
$N_{\sigma}/N_v=\{12.6,\,8.7,\,5.5\}$.
An important question beyond the scope of the present work is to elucidate the
relationship (if any) between the vortex array melting observed here and any
impending transition into a quantum Hall state (not accessible with the
current formalism). Fractions $N/N_v\lesssim 6$ have been predicted to favor a 
quantum Hall state in Bose gases~\cite{Cooper}.

\begin{figure}[t]
\epsfig{file=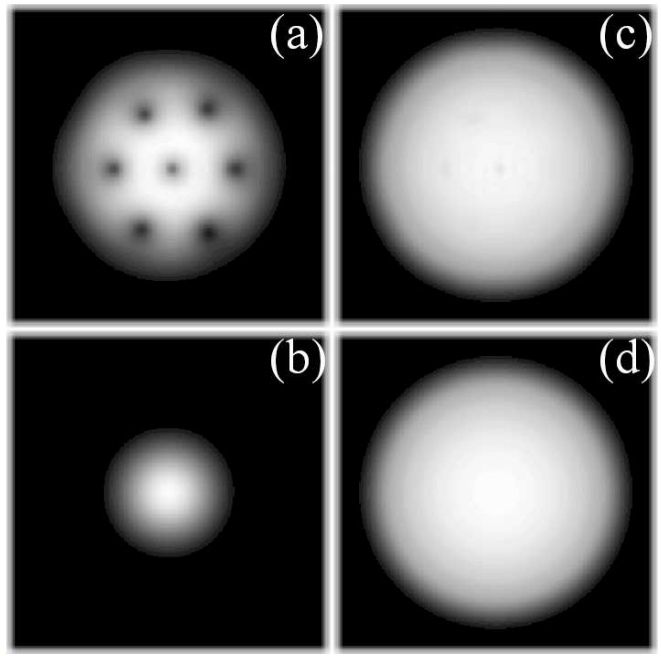,width=\columnwidth,angle=270}
\caption{Density plots for the in-trap gap function $\Delta({\bf x})$ (a,b) 
and number density $n_{\sigma}({\bf x})$ (c,d) for $\alpha=40$ and 
$\Omega/\omega_{\rho}=0.4$, for which $k_BT_c/\hbar\omega_{\rho}\approx 2.03$. 
Figures (a,c) and (b,d) correspond to $k_BT/\hbar\omega_{\rho}=1$ and 2, 
respectively [the $T=0$ cases are shown in Fig~\ref{fig2}(a,d)]. Each box is 
16~$\mu$m on a side.}
\label{fig3}
\end{figure}

Perhaps the most important result of this work is that the vortex cores are
faint but nevertheless clearly visible as depressions in the particle density,
particularly for lower $\Omega$ and $T$. For $\Omega=0.4$ and $0.7$, the 
density in the vortex core is reduced by approximately $1/2$ and $1/4$, 
respectively. Such large depressions were previously thought only to occur far 
from the weak-coupling BCS limit relevant to these calculations~\cite{Bulgac2}. 

As shown in Fig.~\ref{fig3}, the results for a given $\Omega$ are strongly
affected by the temperature. As $T$ increases, only atoms in the vicinity of 
the trap center participate in the pairing because the local value of $T_c$ is 
lowest at the low-density surface of the cloud; alternatively, most of the 
pair-breaking quasiparticle excitations occur at the surface where the 
effective potential is lowest and the tangential particle velocities are 
largest. Similar results have been obtained with a local-density 
Ginzburg-Landau theory~\cite{Rodriguez}. Because the healing length, which 
governs the vortex core size, also diverges as the temperature increases, 
fewer vortices can fit into the reduced superfluid region. The visibility of 
vortices in the particle density is reduced at finite temperature due to the 
thermal occupation of core states.

The most important outstanding issue to be addressed in future work is the 
vortex coherence between the wells of the optical lattice. For deep lattices
such as are considered here the Josephson coupling between sites becomes
small~\cite{Martikainen}. In this regime, vortex lines could break up into 
disconnected `pancake' vortices in each layer, rendering them unobservable 
experimentally. In the high-$T_c$ superconductors, this vortex decoupling
occurs when the anisotropy parameter $\gamma=\sqrt{m_z/m_{\rho}}
\gtrsim 150$~\cite{Clem,Tyagi} ($m_z$ and $m_{\rho}$ are 
the effective masses perpendicular and parallel to the layers), though a full 
description of such a decoupling transition is not yet complete~\cite{Zamora}. 
Assuming $m_z=\hbar^2/\left(\partial^2\varepsilon_k/\partial k^2\right)$ where 
$\varepsilon_k$ is the dispersion of the lowest-lying band near the trap center,
the decoupling $\gamma$ corresponds to an optical lattice depth approaching
$70E_R$. This value is much lower and higher than those of 
Refs.~\onlinecite{OHara1} and \onlinecite{Zwierlein2}, respectively, suggesting
that the melting of well-defined vortex lines should be directly observable 
in experiments employing Nd:YAG (but not CO$_2$) lasers. A full calculation 
incorporating interwell tunneling remains necessary; the disappearance of 
vortices observed experimentally could be due not only to vortex decoupling but 
also to the onset of a quantum Hall phase, which is expected to restore the 
cylindrical symmetry broken by the vortices.

\begin{acknowledgments}
It is a pleasure to thank G.~M.~Bruun, Z.~Dutton, J.-P.~Martikainen, and 
N.~Nygaard for stimulating discussions. This work was supported in part by
the Canada Foundation for Innovation and NSERC.
\end{acknowledgments}


\begin{thebibliography}{99}
\bibitem{comment}
For the parameters used in the present calculations, the density dependent
correction to the 2D coupling strength~\cite{Petrov1} is on the order of 10\%, 
and can be neglected.
\end{thebibliography}
\end{document}